\documentclass[12pt]{iopart}
\usepackage{iopams}
\usepackage[utf8]{inputenc}
\usepackage{hyperref}
\usepackage{graphicx}
\expandafter\let\csname equation*\endcsname\relax 
\expandafter\let\csname endequation*\endcsname\relax 
\usepackage{amsmath, amsfonts}
\usepackage{amssymb}
\usepackage{float}

\usepackage{xcolor, soulutf8}

\usepackage{caption}
\usepackage{multirow}
\begin{document}

\title[Barrow HDE in Brane]{Barrow Holographic Dark Energy in Brane World Cosmology}

\author{Anirban Chanda$^1$}
\ead{anirbanchanda93@gmail.com}
\author{Arpan Krishna Mitra$^2$}
\ead{arpankmitra@aries.res.in}
\author{Sagar Dey$^1$}
\ead{sagardey231@gmail.com}
\author{Souvik Ghose$^{3}$ \footnote{Author to whom any correspondence should be addressed.}}
\ead{dr.souvikghose@gmail.com}

\author{B. C. Paul$^1$}
\ead{bcpaul@associates.iucaa.in}
\address{$^1$Department of Physics, University of North Bengal, 
Rajarammohunpur 734105}
\address{$^2$Aryabhatta Research Institute of Observational Sciences (ARIES), Nainitaal, 263001}
\address{$^3$Harish-Chandra Research Institute, Prayagraj, 211019}

\vspace{10pt}
\begin{indented}
\item[]August 2017
\end{indented}

\begin{abstract}
Cosmological features of Barrow Holographic Dark Energy (BHDE), a recent generalization of original Holographic dark energy with a richer structure, are studied in the context of DGP brane, RS II brane-world, and the cyclic universe. It is found that a flat FRW scenario with pressure less dust and a dark energy component described as BHDE can accommodate late time acceleration with Hubble horizon considered as infrared cut off even in the absence of  interaction between the dark sectors. Statefinder diagnostic reveals that these model resemble $\Lambda CDM$ cosmology in future. It is found that  BHDE parameter $\Delta$, despite its theoretically constrained range of values, is significant in describing the evolution of the universe, however,  a classically stable cosmological model cannot be obtained in the RS-II and DGP brane. Viability of the models is also probed with observed Hubble data.
\end{abstract}

%\maketitle

\section{Introduction}

Cosmological observations  predict that the universe is undergoing an accelerated phase of expansion \cite{riess1998fmf, pmutter:1998vns, Bengaly:2019ibu}. One way to account for this late-time acceleration is by introducing a constant term, often referred to as the cosmological constant, into the equations of General Relativity. However, the observed value of this constant appears to be exceedingly small and cannot be naturally explained within the framework of fundamental physics, primarily due to the lack of a complete quantum theory of gravity.  Within the scientific literature, three distinct approaches have been outlined to address this cosmic conundrum.
In the first approach, the Einstein's theory of  gravitation is modified, so  that at a large scale it would produce the correct phenomenology and would reduce to  GR at the low energy limit. Efforts in this direction have lead to formulation of theories namely, $f(R)$ gravity \cite{de2010f, nojiri2011unified, Fabris:2023opv, DOnofrio:2023jgc, Odintsov:2023cli}, $f(G)$ gravity \cite{nojiri2005modified, Twagirayezu:2023owr}, $f(R, G)$ gravity \cite{Saha:2023hyc}, $f(T)$ gravity \cite{chen2011cosmological, linder2010einstein, Bhaumik:2021kcm} and many others. The second approach is to leave GR unchanged as the underlying theory of gravity but modify the matter sector by introducing the scalar fields or the exotic fluids in the Einstein's field equation. When describing the late time acceleration, these dynamical models come under the umbrella term 'dark energy'. There is also a third approach that modifies the underlying theory of gravity but in a very different manner and is sometimes motivated by the unification of gravity with other gauge forces. These theories often introduce extra dimensions, yielding an effective four-dimensional theory as a limiting case. The origin of the higher dimensional theories of gravity  goes back to the time when  Kaluza \cite{kaluza2018unification} and Klein \cite{klein1926quantentheorie} proposed to unify gravity with electromagnetic interactions by introducing a fifth dimension. 
The extra dimension required in these theories is compactified to avoid any inconsistency arising in the four-dimensional world we experience. 
Brane-world scenarios are also obtained proceeding in this third direction but unlike the KK theory, the commodification of the extra dimensions is not needed in these frameworks. These theories originated as phenomenological models at first and were later revived in the framework of string theory. The basic idea is that our four-dimensional universe exists as a brane (hypersurface) in a higher dimensional bulk. While gravity can propagate through this bulk all the other forces are confined on the brane only. The dark energy is often realized as bulk induced effect in these models.  From a cosmological point of view, there are three major schemes  discussed in the literature in this direction: (i) Dvali-Gabadadze-Porrati (DGP hereafter) brane-world \cite{dvali20004d} , (ii) Randall and Sundrum (RS II) brane-world \cite{randall1999large, randall1999alternative}, and (iii) the cyclic universe proposed by Steinhardt and Turok \cite{steinhardt2002cosmic}.
In DGP brane-world scenario, a mechanism is suggested where 4-D Newtonian gravity emerges on a 3-brane sitting in a 5D Minkowski bulk with an extra dimension with infinite size. There are two main branches of the DGP brane cosmology. There is a self-accelerating branch that does not require dark energy to generate late-time acceleration. However, this branch is plagued with the emergence of ghost degrees of freedom\cite{Luty:2003vm, Charmousis:2006pn}. Moreover, the predicted cosmological parameters in this branch conflicts with the recent observations \cite{Fairbairn:2005ue, Maartens:2006yt, Fang:2008kc}. The other branch, often known as the normal branch, requires at least a cosmological constant to generate late time acceleration otherwise it is free from the problems faced by the former model.  Randall and Sundrum approach is much older than that of DGP brane formalism. Unlike a Minkowski bulk in DGP models,  Randall and Sundrum \cite{randall1999alternative} considered a non-flat, wrapped bulk geometry. The bulk spacetime in the model contains a negative cosmological constant. The curvature of bulk was shown to induce a Newtonian theory of gravity on the brane. 
In the present work,  RS II brane model is considered to explore the cosmological evolution with Barrow holographic Dark energy (BHDE) \cite{barrow2020area}.
The cyclic universe was originally conceived from M-theory. According to this model the universe  undergoes an endless sequence of epochs, always starting from a big bang and ending with a so-called big-crunch \cite{steinhardt2002cosmic}.  This idea was incorporated in the bouncing brane-world scenario \cite{shtanov2003bouncing}. Here too, a five-dimensional cosmological constant in the bulk was necessary to induce its four-dimensional analogue to the four-dimensional brane universe. 
Most of the promising cosmological models in the brane-world framework thus also require the presence of a dark energy component, at least in the form of a cosmological constant. Holographic consideration in cosmology provides an intriguing approach to the description of dark energy \cite{hooft1993dimensional, susskind1995world, bousso2002holographic}. If dark energy is a manifestation of vacuum energy, at a cosmological scale Holographic Dark Energy (HDE hereafter) might be well suited to describe it \cite{cohen1999effective, hovrava2000probable}. HDE has indeed proven to be efficient in describing the late time acceleration \cite{li2004model, wang2017holographic, horvat2004holography, pavon2005holographic, wang2005transition, setare2009non, setare2006interacting} and is in good agreement with observational data \cite{zhang2005constraints, feng2007testing, li2009holographic, zhang2009holographic, d2019holographic, sadri2019observational}. Many HDE models have been proposed over the years mainly due to the lack of agreement over which horizon should be considered as the largest distance of the model. Also, new HDE models have appeared which use the standard holographic principle but  are based on more generalized entropy instead of the original Bekenstein-Hawking one \cite{Nojiri:2022sfd, Nojiri:2022dkr, Nojiri:2022ljp, Nojiri:2021jxf, Liu:2021heo, Pandey:2021fvr, Sharma:2021zjx, Moradpour:2020dfm, Ghaffari:2019mrp, Ghaffari:2019itm, Sharma:2020awh}. 
A plethora of study exists in the literature on differnt aspects of HDE in brane models (e.g. \cite{Farajollahi:2014hzp, Astashenok:2019rwt, Belkacemi:2018dgy}). Recently, Ghaffari {\it et. al.}  have studied one such HDE model, namely Tsallis HDE (THDE hereafter) in brane-world scenario \cite{ghaffari2019tsallis}. The Friedman equations for THDE and BHDE seem similar when the Hubble Horizon is chosen as the infrared cutoff. However, the physical motivation behind these two models are entirely different. THDE is obtained by generalizing Boltzmann–Gibbs additive entropy to non-additive entropy. On the other hand, the origin of BHDE lies in the quantum gravity induced intricate fractal structure of the Black-Hole horizon. There is a lack of consensus on the range of THDE parameter ($\delta$) in the literature apart from the fact that for $\delta = 1$ THDE reduces to standard HDE. For example, in \cite{ghaffari2019tsallis} $\delta>2$ is used in many instances where as in \cite{Sheykhi:2022gzb}  THDE parameter ($\beta$ in the paper) is argued to be $< 2$. Observational constraints on the THDE parameter \cite{sadri2019observational} while considering Hubble horizon as cut off is $2.211^{+0.121}_{-.181}$. The BHDE parameter, on the other hand, $\Delta$ has to satisfy a theoretical bound $0 \leq \Delta \leq 1$. A recent study \cite{Sheykhi:2022gzb} also shows considerable deviation in the growth of perturbation in THDE and BHDE despite the similarity of the modified Friedman equations in both the cases. As suggested in \cite{Sheykhi:2022gzb}, this dissimilarity probably originates from the difference in the theoretically allowed range of THDE and BHDE. Despite the apparent similarity of the field equations, two significant reasons underlie the current study. Firstly, it is worth considering BHDE in braneworld and explore if satisfactory evolution of relevant cosmological parameters can still be realised within BHDE while satifying the stringent theoretical constraint already imposed on $\Delta$. The Tsallis parameter $\delta$ and the Barrow parameter $\Delta$ are not related through any scaling. The theoretical constraint on the Barrow parameter make this exploration non-trivial (as we shall show). Secondly, although there is no theoretical constraint on the Tsallis parameter ($\delta$), observation would put some bound on it depending on the model. In the case of the BHDE model, the parameter has to satisfy both the theoretical and observational bound. Although a full-scale data analysis is beyond the scope of the present paper, we test both the THDE and BHDE model with Observed Hubble Data (OHD hereafter) \cite{Sharov:2018yvz} and compare our findings. Our findings suggest a clear advantage for BHDE over THDE in some frameworks.
In the present paper, we study the importance of the recently proposed Barrow HDE (BHDE here onward) in DGP brane, RS II brane, and the cyclic universe. In the next section, we present a brief overview of BHDE and outline our motivation for the work. In sec. (\ref{sec:dgp}), (\ref{sec:rs}), and (\ref{sec:cyclic}) we analyze the cosmological parameters, namely the dark energy density parameter($\Omega_D$), dark energy equation of state (EOS hereafter) parameter ($\omega_D$), and deceleration parameter$q(z)$). Then in sec. (\ref{sec:stf}) we elaborate the statefinder diagnostic and discuss the dynamical nature of the dark energy as it evolved. The classical stability of these models is discussed in sec. (\ref{sec:stable}). In sec. (\ref{sec:comp}) we compare the earlier THDE model and the present BHDE model in light of OHD. We finally summarize our findings in sec. ($\ref{sec:concl}$).

\section{Brief overview of BHDE}
Recently, Barrow has suggested a modification of Bekenstein - Hawking (BH here onwards) entropy as quantum deformation may lead to the black hole horizon acquiring a fractal structure which effectively increases its surface area \cite{barrow2020area}. Barrow started from the Schwarzschild black hole and went on to produce what he termed as the rough black hole by attaching many smaller spheres to touch its outer surface and yet smaller spheres touching the surfaces of those spheres. This process is continued to obtain a fractal structure of the horizon. The modified entropy of this rough black hole is then given by:
\begin{equation}
\label{eq:bentropy}
S_B = \left(\frac{A}{A_0}\right)^{1 + \Delta / 2},
\end{equation}
where $A$ is the standard horizon area, $A_0$ is the Planck area and $\Delta$ is the quantifier for the quantum deformation. $\Delta = 0$ corresponds to no deformation and $\Delta = 1$ implies maximum deformation. While the standard HDE is realized from the inequality $\rho_{DE}\leq L^4$, as shown in \cite{saridakis2020barrow}, the use of Barrow entropy instead of BH entropy leads to:
\begin{equation}
\label{eq:bhde}
\rho_{DE} = C L^{\Delta - 2},
\end{equation}
where $S \propto A \propto L^2$. Here, $L$ is the horizon length and $C$ is a parameter with dimension $[L]^{-2 - \Delta}$. BHDE has been studied in different frameworks \cite{paul2022bianchi, saleem2022exact}. In the following sections, we study BHDE in a different scenario considering Hubble horizon as the infrared cut-off. As WMAP and other observations favour a flat universe we restrict ourselves to FRW type models with $k = 0$. In the standard GR framework, BHDE has been constrained from cosmological observations in \cite{sardi2020}. The mean value for $\Delta$ is found to be $0.094$. As $\Delta \rightarrow 0$ corresponds to the original HDE model, BHDE seems to deviate only marginally, the maximum deviation situation being $\Delta \rightarrow 1.0$. We are interested to see here how the evolution of the cosmological parameters depends on $\Delta$ and whether the acceptable nature of evolution can be realized with BHDE deviating only a little from the normal HDE models. To achieve this goal we set up the governing equations in each case and numerically solve them. Relevant cosmological parameters are then plotted against redshift for different $\Delta$ values. Whenever possible we choose these values to span the entire theoretically allowed range for $\Delta$ (i.e. $0 < \Delta < 1$). It is seen that interesting comments can be made on the values of $\Delta$ in each of the frameworks under discussion.  
%--------------------------------- DGP brane section ------------------------------------------
\section{BHDE in DGP Brane}

\label{sec:dgp}
\begin{figure}[H]
	\centering
	\includegraphics[width = 0.8\linewidth]{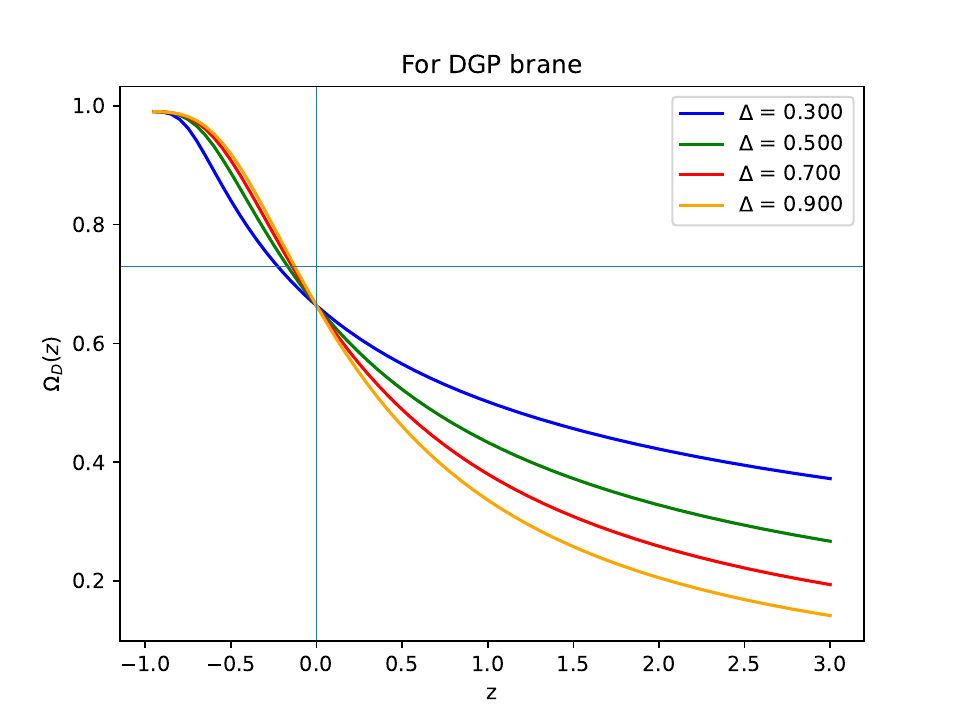}
	\caption{Evolution of DE density in DGP brane. We have considered $\Omega_{D0} = 0.73$ $\epsilon = -1$ and $\Omega_{r_c} = 0.002 \cite{ghaffari2019tsallis, Xu:2013ega}$ }
	\label{fig:omd_dgp}
\end{figure}

\begin{figure}[H]
	\centering
	\includegraphics[width = 0.8\linewidth]{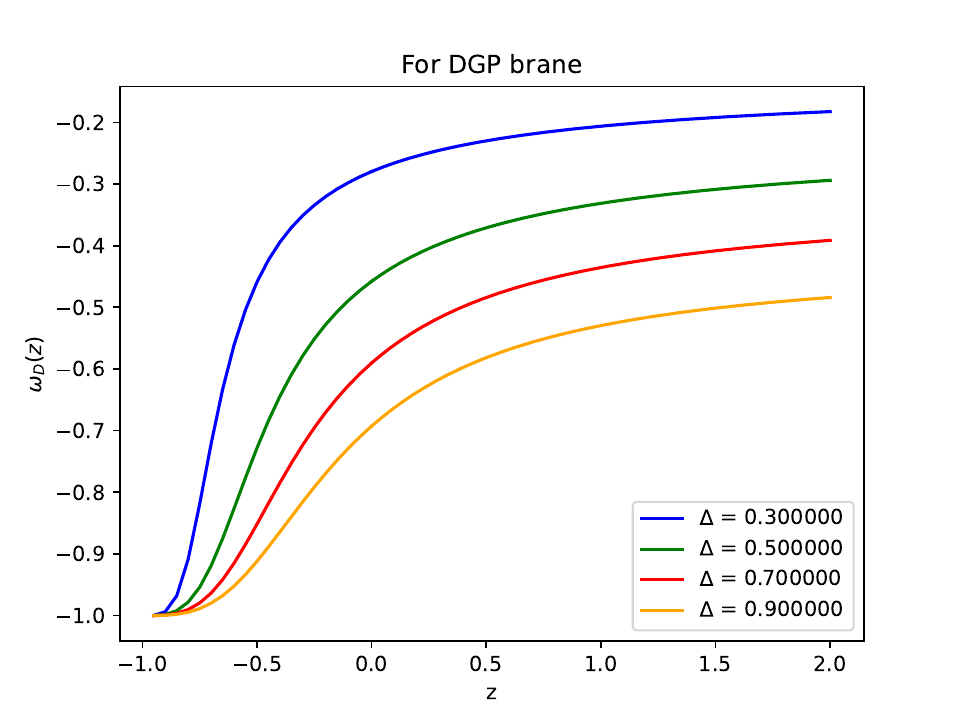}
	\caption{Evolution of dark energy EOS in DGP brane. We have taken $\epsilon = -1$ and $\Omega_{r_c} = 0.002 \cite{ghaffari2019tsallis, Xu:2013ega}$. }
	\label{fig:eos_dgp}
\end{figure}

\begin{figure}[H]
	\centering
	\includegraphics[width = 0.8\linewidth]{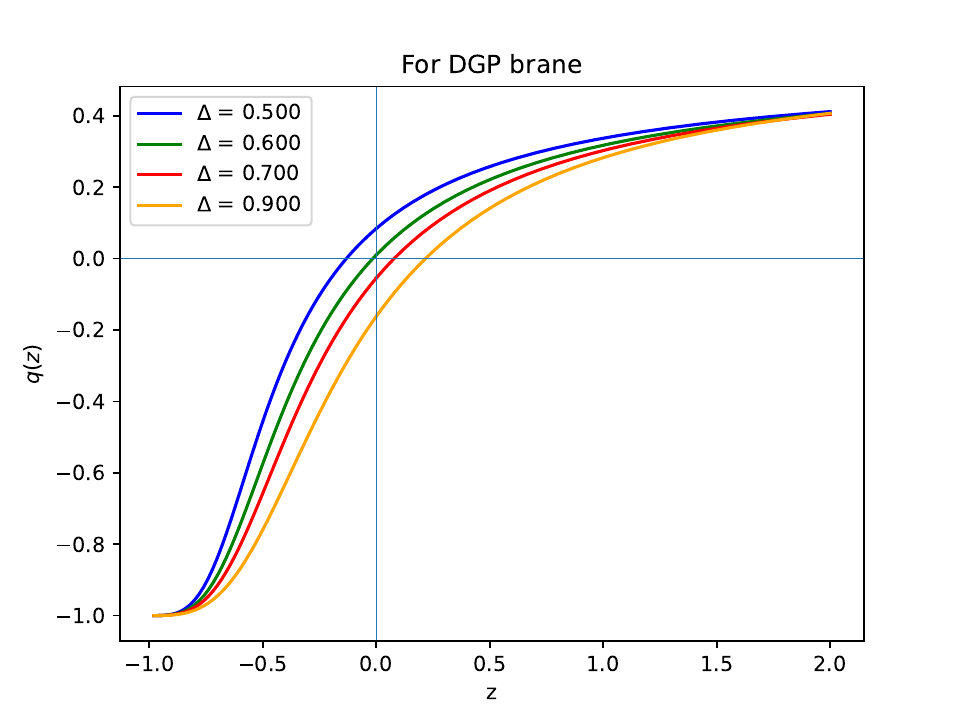}
	\caption{Evolution of deceleration parameter in DGP brane, considering  $\epsilon = -1$ and $\Omega_{r_c} = 0.002 \cite{ghaffari2019tsallis, Xu:2013ega}$. }
	\label{fig:decl_dgp}
\end{figure}

The FRW equation is modified for a flat brane embedded in a five-dimensional bulk \cite{deffayet2001cosmology, ghaffari2019tsallis}:
\begin{equation}
\label{frd-dgp}
H^2 = \left( \sqrt{\frac{\rho}{3M_{pl}^2} + \frac{1}{4r_c^2}}+\frac{\epsilon}{2r_c}\right)^2,
\end{equation}
where we have assumed that $\rho$ is composed of dark matter, normal matter ($\rho_m$), and dark energy on brane ($\rho_D$). Here, $r_c = \frac{M_{pl}^2}{2M_5^3}$ is the crossover scale for DGP brane which separates $4D$ and $5D$ cosmology in the framework, with {\bf $M_{pl}=\sqrt{\frac{1}{8\pi G}}$ being the Plank mass}.  One can still make use of the conservation equations though \cite{ghaffari2019tsallis}. Also, $\epsilon = \pm 1$ specifies the two different branches of DGP brane. For $\epsilon = +1$, one gets a self-accelerating model. The model with $\epsilon = -1$ requires dark energy, at least in form of a cosmological constant \cite{Sheykhi:2015nba} to explain the late time acceleration. We explore the later case throughout the present work as the former is plagued with many theoretical and observational issues. For our model we are considering pressureless matter, which also includes dark matter. We are also considering a dark energy component. Since we are interested in late time universe, such a composition is justified. In the present work we do not consider any interaction between the dark sectors. Hence,
\begin{equation}
\label{dark-consv}
\dot{\rho}_D + 3H(1 + \omega_D)\rho_D = 0,
\end{equation}
and
\begin{equation}
\label{mat-consv}
\dot{\rho}_m + 3H\rho_m = 0 \rightarrow \rho_m = \rho_{m0}(1+z)^3,
\end{equation}
where $\omega_D$ is dark energy equation of state (EoS here after) parameter, $H$ is the Hubble parameter, and $\rho_{m0}$ is the matter density in the present epoch. 
It is convenient to define dimensionless density parameters:
\begin{equation}
\label{den-param}
\Omega_m = \frac{\rho_m}{\rho_{cr}}= \frac{\rho_m}{3M_{pl}^2H^2}, \; and \; \; \Omega_D = \frac{\rho_D}{3M_{pl}^2H^2}
\end{equation}
where $\rho_{cr} = 3M_{pl}^2H^2$ is called the critical density. For $r_c >> 1$, eq. \eqref{frd-dgp} reduces to $H^2 = \frac{\rho}{3M_{pl}^2}$, thus recovering standard Friedmann cosmology in $4D$. Considering only the first order terms in $\frac{1}{r_c}$, as shown in \cite{Sheykhi:2015nba, ghaffari2019tsallis},  eq. \eqref{frd-dgp} can also be recasted as:
\begin{equation}
\label{frw-dgp2}
H^2 - \frac{\epsilon}{r_c}H = \frac{\rho}{3M_{pl}^2}.
\end{equation}
Writing $\rho = \rho_m + \rho_D$ and using eq. \eqref{den-param}, eq. \eqref{frw-dgp2} becomes:
\begin{equation}
\label{den-dgp}
\Omega_m + \Omega_D + 2 \epsilon \sqrt{\Omega_{r_c}} = 1,
\end{equation}
where we have defined $\Omega_{r_c} = \frac{1}{4H^2r_c^2}$. Now, describing the dark energy part as BHDE (eq. \eqref{eq:bhde}) and using Hubble horizon as the IR cutoff of the theory one obtains,
\begin{equation}
\label{Omd-dgp}
\Omega_D = \frac{CH^{-\Delta}}{3M_{pl}^2}.
\end{equation}
Now, using the time derivative of eq. \eqref{eq:bhde} with $L = H^{-1}$ and eq. \eqref{Omd-dgp} one gets
\begin{equation}
\label{omprime-dgp}
\Omega_D^{\prime} = - \Delta \Omega_D \frac{\dot{H}}{H},
\end{equation}
where prime stands for derivative with respect to $ln(a)$, $a$ being the scale factor, and $\dot{\Omega}_D = H \Omega_D^{\prime}$. Then from eqs. \eqref{eq:bhde}, \eqref{frw-dgp2} and \eqref{den-param} one gets
\begin{equation}
\label{hdot-dgp}
\frac{\dot{H}}{H^2} = \frac{- 3 (1 - \Omega_D - 2\epsilon \sqrt{\Omega_{r_c}})}{(\Delta - 2 )\Omega_D- 2\epsilon \sqrt{\Omega_{r_c}} +2},
\end{equation}
which on substitution gives from eq. \eqref{omprime-dgp}
\begin{equation}
\label{omprime-dgp2}
\Omega_D^{\prime} = \frac{3 \Delta \Omega_D (1 - \Omega_D - 2\epsilon \sqrt{\Omega_{r_c}})}{(\Delta - 2 )\Omega_D- 2\epsilon \sqrt{\Omega_{r_c}} +2}.
\end{equation}
We solve this eq. \eqref{omprime-dgp2} numerically to obtain time evolution of $\Omega_D$ which is shown in fig. (\ref{fig:omd_dgp}). For larger $\Delta$ values dark energy density is found to be evolving from a much smaller value in the earlier epoch compared to that with smaller $\Delta$ values where dark energy density is significantly higher in earlier epochs. As $\Delta$ is decreased beyond a certain limit ($\Delta\sim 0.15$) the density parameter, $\Omega_{D}$, exhibits a monotonically increasing behaviour with values approaching $1$ in the present time and exceeding $\Omega_{D}=1$ in the near future. Thus, viable cosmological models can not be obtained in this case below a certain limiting value of $\Delta$.
%------------------ Density in DGP------------------------------------
%-------------------------------------------------------------------------
Expression for EoS parameter $\omega_d$ and the deceleration parameter $q$ can be obtained by some algebra, which are given by
\begin{equation}
\label{eos-dgp}
\omega_D = - 1 + \frac{(\Delta - 2)(\Omega_D + 2\epsilon \sqrt{\Omega_{r_c}} -1)}{(\Delta -2)\Omega_D - 2\epsilon \sqrt{\Omega_{r_c}} +2},
\end{equation}
and
\begin{equation}
\label{decl-dgp}
q = -1 + \frac{3 (1 -\Omega_D - 2\epsilon \sqrt{\Omega_{r_c}})}{(\Delta -2)\Omega_D - 2\epsilon \sqrt{\Omega_{r_c}} + 2}.
\end{equation}
The evolution of the EoS parameter (eq. \eqref{eos-dgp}) and the deceleration parameter (eq. \eqref{decl-dgp}) are shown in fig. (\ref{fig:eos_dgp}) and (\ref{fig:decl_dgp}) respectively.

%------------------------- EOS and q in DGP ------------------------------------------------

Evidently (\ref{fig:eos_dgp}), dark energy never becomes phantom and approaches $\Lambda$CDM values in future for any $\Delta$. BHDE is able to produce late time acceleration. The deceleration to acceleration transition point varies with $\Delta$. For smaller values of $\Delta$ (for example, $\Delta \leq 0.6$), this transition happens in some future epoch. We have checked that for yet smaller $\Delta$ values (e.g. $\Delta = 0.1$) the evolution becomes nonphysical as $\Omega_D$ diverges at around $z = 1$.

%%%%%%%%%% RS II %%%%%%%%%%%%%%%

\section{BHDE in RS II Brane}
To explore BHDE in RS II brane-world framework, we start with the modified Friedmann equation on the brane \cite{ghaffari2019tsallis}.

\label{sec:rs}
\begin{equation}
\label{hubrs}
H^{2} + \frac{k}{a^{2}}= \frac{8\pi}{3M_{pl}^{2}}(\rho_{m} + \rho_{\Lambda}),
\end{equation}
where, $\rho_{m}$ is  the energy density of the pressureless source, and,
\begin{equation}
\label{rholam}
\rho_{\Lambda}=\rho_{\Lambda 4}=\frac{M_{pl}^{2}\rho_{\Lambda 5}}{32\pi M_{5}^{3}}+ \frac{3M_{pl}^{3}}{2\pi\left( \frac{L_{5}}{8\pi}-2r_{c}^{2}\right)}
\end{equation}
is the four dimensional effective DE density on the brane \cite{ghaffari2019tsallis}.

Here, $\rho_{\Lambda 5}$ is the $5D$ bulk holographic DE, which, in case of BHDE takes the following form, $$\rho_{\Lambda 5}=\frac{3c^{2}M_{5}^{3}}{4\pi}C H^{2-\Delta}$$.
When combined with \eqref{rholam}, we get effective 4D BHDE density
\begin{equation}
\label{modrho}
\rho_{\Lambda}=\rho_{\Lambda 4}=\frac{3M_{pl}^{2}c^{2}C H^{2-\Delta}}{128\pi^{2}}+ \frac{3M_{pl}^{3}}{2\pi\left( \frac{L_{5}}{8\pi}-2r_{c}^{2}\right)}.
\end{equation}

Since, we are interested in the late time behaviour of the universe, the horizon is large enough to simplify the above equation \cite{ghaffari2019tsallis}. Thus, for large values of $L$ we can drop the second term and obtain,
\begin{equation}
\label{modrh1}
\rho_{\Lambda}=\frac{3M_{pl}^{2}c^{2}C H^{2-\Delta}}{128\pi^{2}}.
\end{equation} 
Taking time derivative of \eqref{modrh1} we obtain

\begin{equation}
\label{dorolrs}
{\dot{\rho}_{\Lambda}}=(2-\Delta)\frac{\dot{H}}{H}\rho_{\Lambda}.
\end{equation}
The relation between density parameter $\Omega$ and energy density $\rho$ has been given in \eqref{den-param}. {\bf One can now recast \eqref{hubrs} in terms of density parameters in the following form,}
\begin{equation}
\label{omers}
1+\Omega_{k}=\Omega_{m}+ \Omega_{\Lambda}
\end{equation}
with $\Omega_{k}=\frac{k}{H^{2}a^{2}}$.

\begin{figure}[H]
	\centering
	\includegraphics[width = 0.8\linewidth]{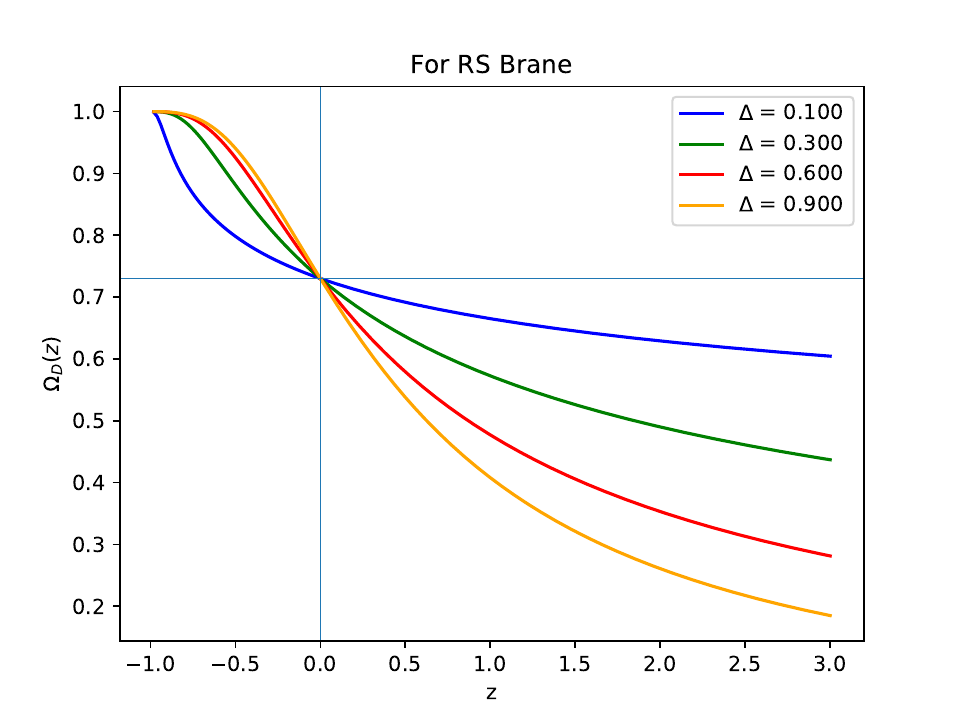}
	\caption{Evolution of DE density in RS II brane. Here, we have taken $\Omega_{D0} = 0.73$. }
	\label{fig:omd_rs}
\end{figure}

On-wards we will concentrate on flat FRW universe as indicated in \cite{ghaffari2019tsallis}. This effectively means $k=0$. 
Time derivative of \eqref{hubrs} when coupled with \eqref{modrh1} and \eqref{dorolrs}, yields,
\begin{equation}
\label{hdothsq}
\frac{\dot{H}}{H^{2}}= \frac{3(\Omega_{\Lambda}-1)}{2+(\Delta -2)\Omega_{\Lambda}}.
\end{equation}
Which can be used to write,
\begin{equation}
\label{omprirs}
\Omega_{\Lambda}'= \frac{\dot{\Omega}_{\Lambda}}{H}= -\Delta\Omega_{\Lambda}\frac{\dot{H}}{H^{2}}= \frac{3\Delta\Omega_{\Lambda}(1-\Omega_{\Lambda})}{2+(\Delta -2)\Omega_{\Lambda}}.
\end{equation}
Numerical solution now gives $\Omega_D = \Omega_{\Lambda}$ at every epoch ($z$). Fig. (\ref{fig:omd_rs}) depicts the evolution of $\Omega_D (\Omega_\Lambda)$ with redshift. In RS II brane scenario too, a lower bound of $\Delta$ exists, beyond which $\Omega_{\Lambda}$ becomes a monotonically increasing function of $z$ and does not converge into . We estimate this bound of $\Delta$ to be around $\Delta\sim 0.1$. Beyond this limit stable cosmological models can not be obtained.

For any $\Delta$ value, the present estimate for the dark energy density ($\approx 0.73$) can be satisfied. This density was far lower in earlier epochs except for the models with small delta values (e.g., $\Delta = 0.1$) where dark energy density decreases very slowly in the past. Dark energy accounts for almost the entire budget in future irrespective of $\Delta$. A few more steps of non-trivial algebra leads us to the expression of EoS and parameter $\omega _{\Lambda}$ of BHDE and deceleration parameter $q$.

\begin{equation}
\label{eosrs1}
\omega_{\Lambda} =-1 +(\Delta -2)\frac{(\Omega_{\Lambda}-1)}{2+(\Delta -2)\Omega_{\Lambda}},
\end{equation}
and
\begin{equation}
\label{dcprs}
q=-1 -\frac{\dot{H}}{H^{2}}=-1- \frac{3(\Omega_{\Lambda}-1)}{2+(\Delta -2)\Omega_{\Lambda}}.
\end{equation}
The evolution of these parameters are shown below in figs. (\ref{fig:eos_rs}) and (\ref{fig:decl_rs}). Much like the DGP brane case, the dark energy EoS parameter evolve into $\omega_D (=\Omega_{\Lambda})=-1.0$ in future for any $\Delta$ value but and is never phantom at any stage of evolution. The universe transits from a decelerating phase into an accelerating one. The transition redshift, however, depends on $\Delta$ and for smaller values such as $\Delta \leq 0.360$ this transition occur only in future making the model incapable of describing the present phase of acceleration.
\begin{figure}[H]
	\centering
	\includegraphics[width = 0.8\linewidth]{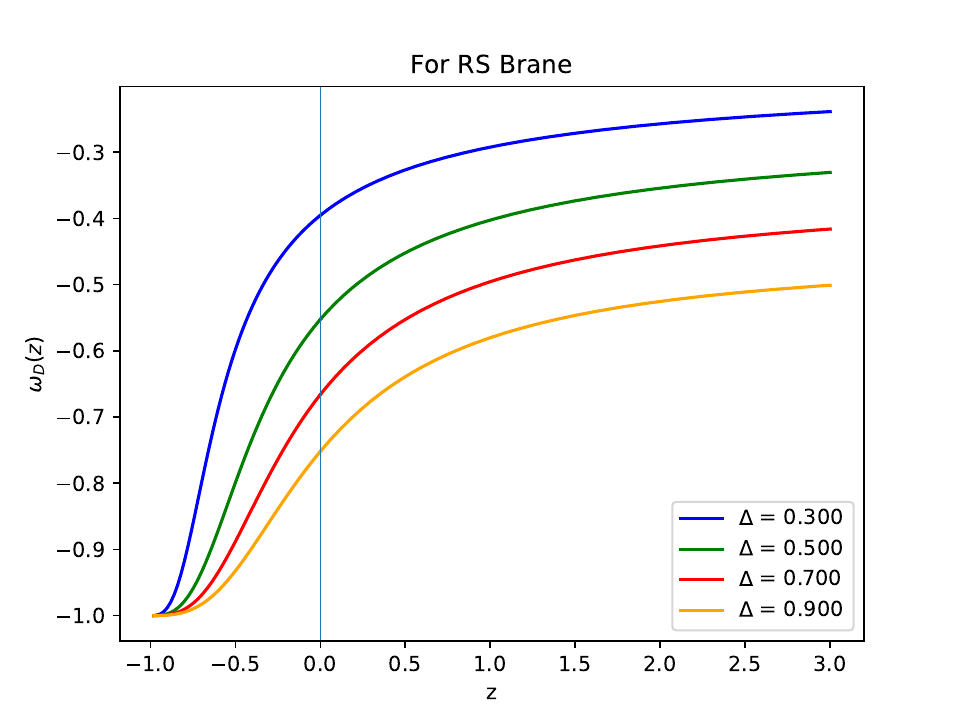}
	\caption{Evolution of dark energy EOS in RS II brane, taking $\Omega_{D0} = 0.73$. }
	\label{fig:eos_rs}
\end{figure}
\begin{figure}[H]
	\centering
	\includegraphics[width = 0.8\linewidth]{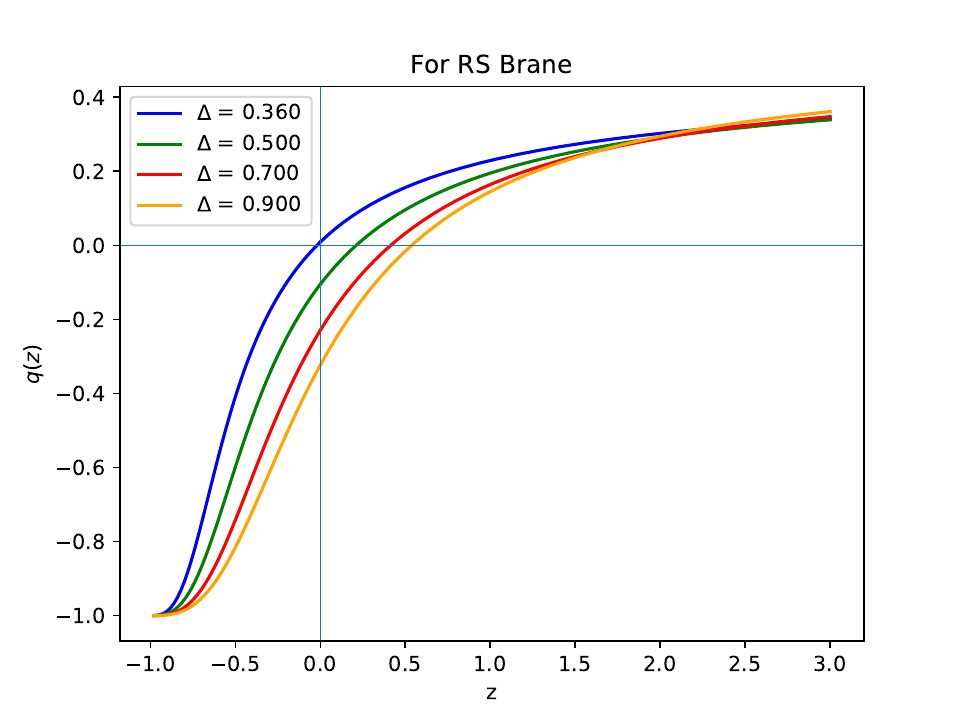}
	\caption{Evolution of deceleration parameter in RS II brane, considering $\Omega_{D0} = 0.73$. }
	\label{fig:decl_rs}
\end{figure}

%%%%%%%%%%%%%%%%%Cyclic%%%%%%%%%%%%
\section{BHDE in Cyclic Universe}
\label{sec:cyclic}
\begin{figure}[H]
	\centering
	\includegraphics[width = 0.8\linewidth]{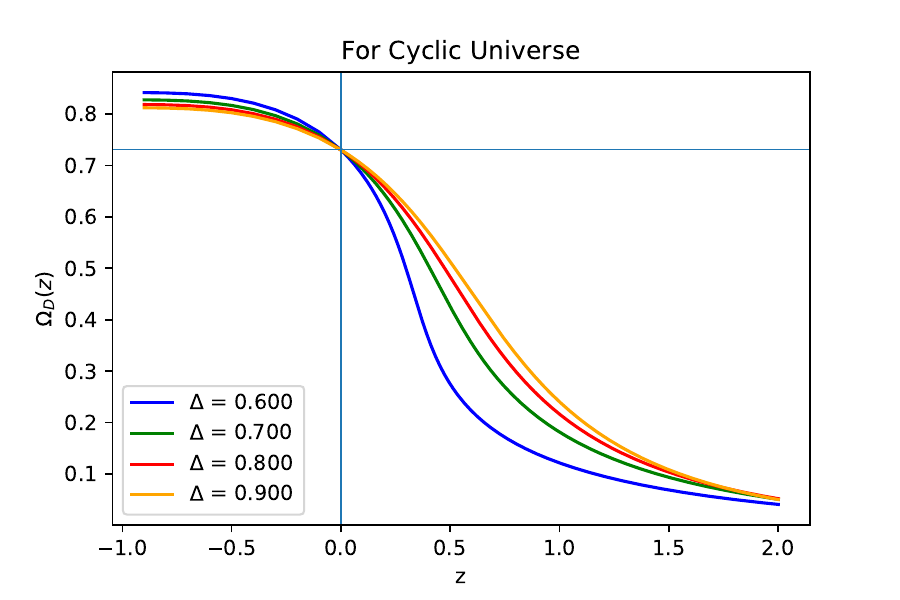}
	\caption{Evolution of DE density in cyclic universe. Here, we have considered $\Omega_{D0} = 0.73$ and $u_0 = 0.3$. }
	\label{fig:omd_cyclic}
\end{figure}

\begin{figure}[H]
	\centering
	\includegraphics[width = 0.8\linewidth]{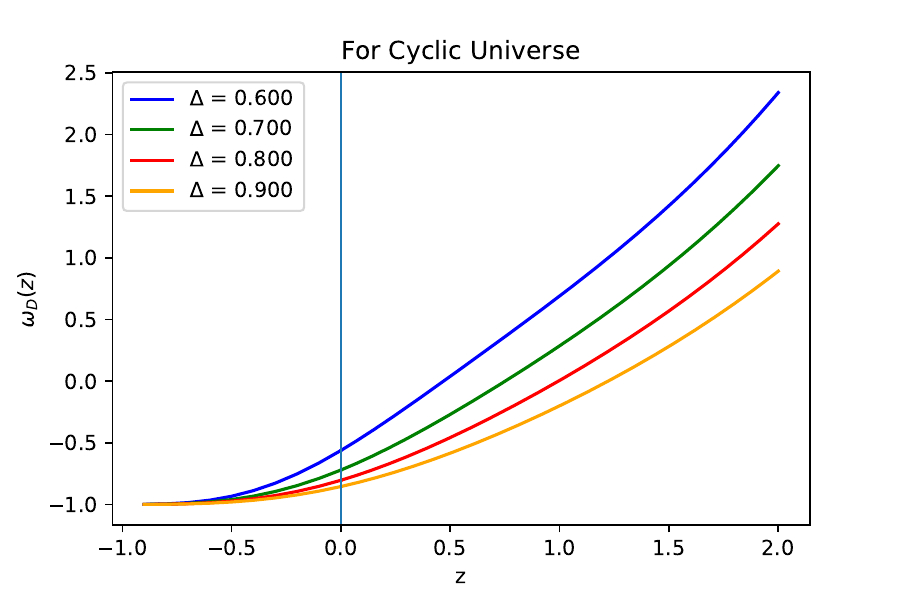}
	\caption{Evolution of dark energy EOS in cyclic universe. Here, we have taken $\Omega_{D0} = 0.73$. }
	\label{fig:eos_cyclic}
\end{figure}

\begin{figure}[H]
	\centering
	\includegraphics[width = 0.8\linewidth]{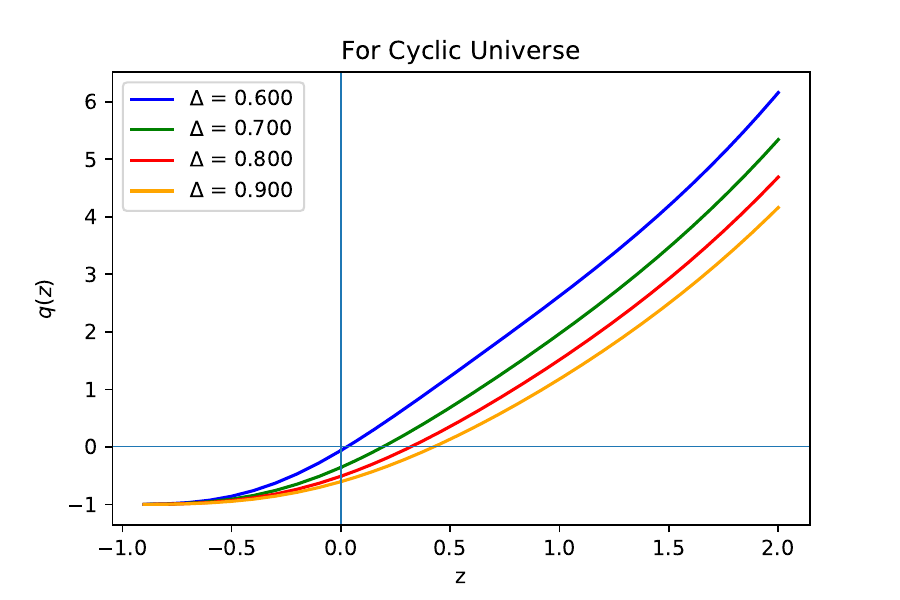}
	\caption{Evolution of deceleration parameter in cyclic universe, considering $\Omega_{D0} = 0.73$. }
	\label{fig:decl_cyclic}
\end{figure}

The modified Friedmann equation in this case is given by \cite{ghaffari2019tsallis, Singh:2006im}:
\begin{equation}\label{cy1}
H^{2}=\frac{\rho}{3M_{pl}^{2}}\Big(1-\frac{\rho}{\rho_{c}}\Big),
\end{equation}
where $\rho$ is the total energy density of the cosmic fluid, and $\rho_{c}$ denotes the critical density constrained by quantum gravity and not to be confused with the usual critical density ($\rho_{cr}=3M_{pl}^{2}H^{2}$). Formally, $\rho_c = \frac{\sqrt{3}}{16 \pi^2 \gamma^3}\rho_{pl}$ (see \cite{Singh:2006im} for detail), where $\rho_{pl} = 1/\hslash G^2$ is the Plank density and $\gamma$ is known as Immirzi parameter. The cosmic fluid in this case is considered to be a combination of dark matter (DM) and dark energy (DE). The different components of the fluid $i.e.$, DM and DE are assumed to be mutually non-interacting so as usual eqs. \eqref{dark-consv} and \eqref{mat-consv} hold as usual on brane.
Putting these and using eq. \eqref{den-param} in eq. \eqref{cy1} we obtain,
\begin{equation}\label{cy5}
\Omega_{D}=(1-\Omega_{m})+\frac{\Omega_{D}+\Omega_{m}}{\frac{\rho_{c}}{\rho_{cr}}-(\Omega_{D}+\Omega_{m})}.
\end{equation}
Using eqs. (\ref{mat-consv}) and (\ref{cy5}) one gets,
\begin{equation}\label{cy6}
\Omega_{D}(z\Rightarrow -1)\approx 1+\frac{\Omega_D}{\frac{\rho_{c}}{\rho_{cr}}-\Omega_{D}}.
\end{equation}
for the limit $z\Rightarrow -1$. From the above equation, it is evident that for $\rho_{c}>\rho_{D}$ we have $\Omega_{D}>1$. For Barrow HDE the DE density parameter can be expressed as,
\begin{equation}\label{cy7}
\Omega_{D}=\frac{C}{3M_{pl}^{2}}H^{-\Delta}.
\end{equation}
We take the time derivative of eq. (\ref{cy1})
and, combine the results with eq. (\ref{den-param})
and eq. (\ref{cy1}) to obtain,
\begin{equation}\label{cy8}
\frac{\dot{H}}{H^{2}}=\frac{-3u(2-\Omega_{D}(1+u))}{2(1+u)+(\Delta-2)(2-\Omega_{D}(1+u))},    \end{equation}
where $u=\frac{\Omega_{m}}{\Omega_{D}}$. The above equation can now be used to obtain the evolution equation of $\Omega_{D}$ as,
\begin{equation}\label{omprime_cycl}
\Omega_{D}'=\frac{3\Delta u\Omega_{D}(2-\Omega_{D}(1+u))}{2(1+u)+(\Delta-2)(2-\Omega_{D}(1+u))},
\end{equation}
where the prime denotes derivative with respect to $ln(a)$. As usual, we solve eq. \eqref{omprime_cycl} numerically and plot the evolution of $\Omega_D$ (see fig. (\ref{fig:omd_cyclic})) for different $\Delta$ values. Dark energy density, as expected, was small in earlier epochs only to become dominant in late time.  
 One can now obtain the expressions for the DE EoS parameter $(\omega_{D})$ and the deceleration parameter ($q$) as,
\begin{equation}\label{eos_cyclic}
\omega_{D}=-1+\frac{u(2-\Delta)(2-\Omega_{D}(1+u))}{2(1+u)+(\Delta-2)(2-\Omega_{D}(1+u))},
\end{equation}
and
\begin{equation}\label{decl_cyclic}
q=-1+\frac{3u(2-\Omega_{D}(1+u))}{2(1+u)+(\Delta-2)(2-\Omega_{D}(1+u))}.
\end{equation}
The evolution of $\omega_D$ and $q$ are shown in fig. (\ref{fig:eos_cyclic}) and (\ref{fig:decl_cyclic}) respectively. For alll values of $\Delta$ the dark energy converges to $\Lambda$CDM type ($\omega_D = -1$) in future. Fig. (\ref{fig:decl_cyclic}) shows that the universe has made a transition from decelerating to accelerating phase at some earlier epoch and the transition redshift is dependent on the Barrow exponent i.e., $\Delta$. The dark energy equation of state parameter obtains a vlalue similar to that of cosmological constant (i.e. $\omega_D \approx -1.0$) in future. This happens for all $\Delta$ values considered.

\section{Statefinder Diagnostic of The Models}
\label{sec:stf}
\begin{figure}[H]
	\centering
	\includegraphics[width = 0.8\linewidth]{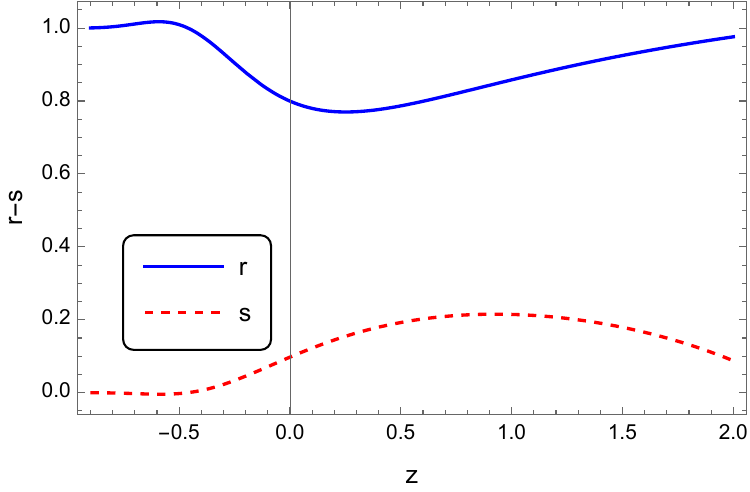}
	\caption{Evolution of statefinder pair in the DGP brane scenario, considering $\Delta = 0.950$. }
	\label{fig:rs_dgp}
\end{figure}
\begin{figure}[H]
	\centering
	\includegraphics[width = 0.8\linewidth]{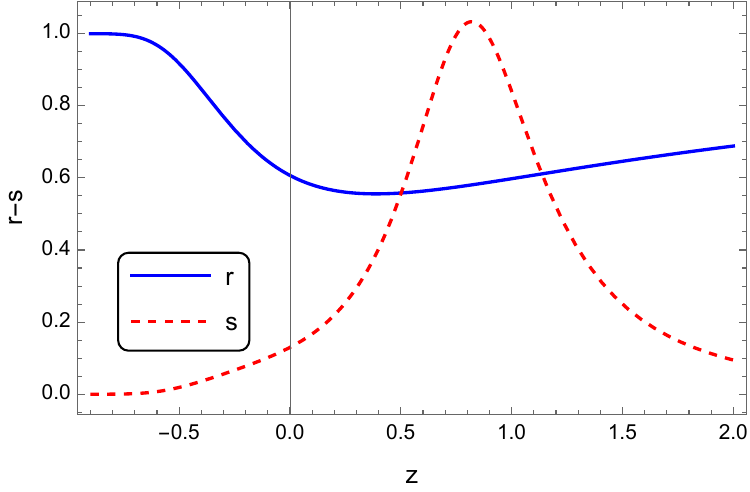}
	\caption{Evolution of statefinder pair in the RSII brane scenario, considering $\Delta = 0.667$. }
	\label{fig:rs_rs2}
\end{figure}
\begin{figure}[H]
	\centering
	\includegraphics[width = 0.8\linewidth]{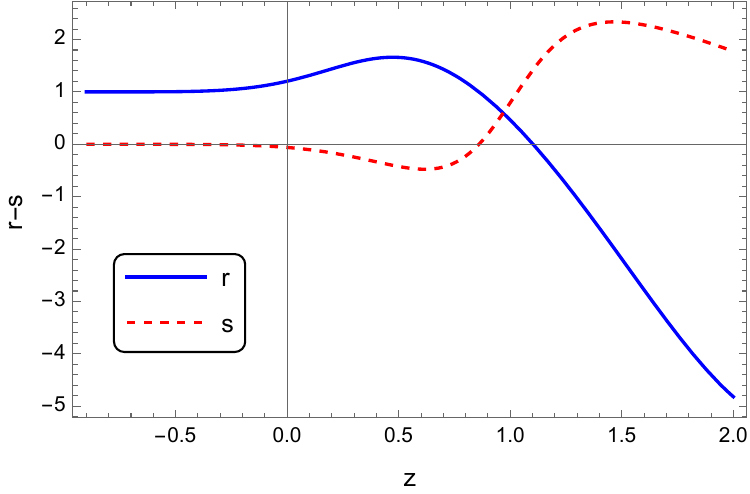}
	\caption{Evolution of statefinder pair in the cyclic universe, considering $\Delta =0.800 $.}
	\label{fig:rs_cyc}
\end{figure}
\begin{figure}[H]
	\centering
	\includegraphics[width = 0.8\linewidth]{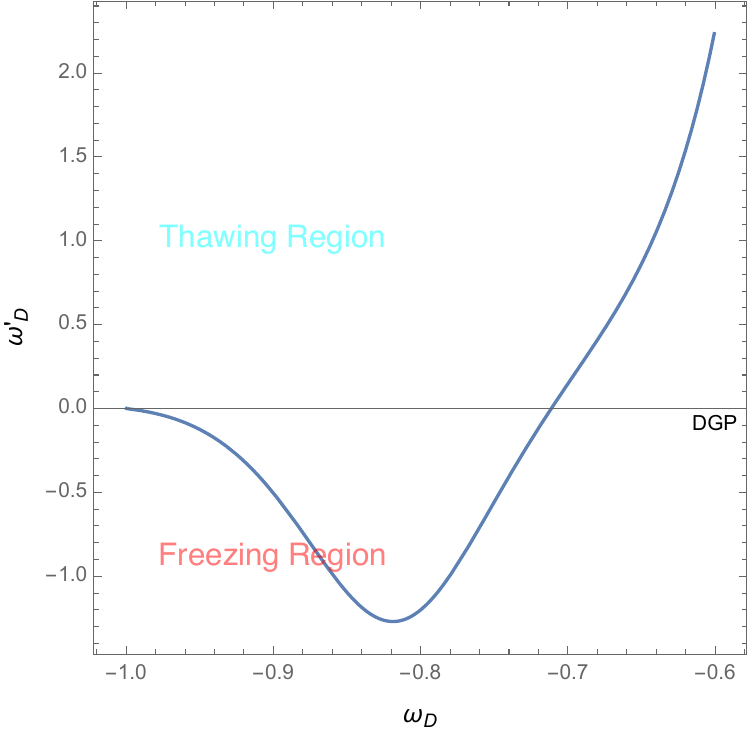}
	\caption{The $\omega^{\prime}_D-\omega_D$ diagram for DGP brane, $\Delta = 0.950$.}
	\label{fig:frz_dgp}
\end{figure}
\begin{figure}[H]
	\centering
	\includegraphics[width = 0.8\linewidth]{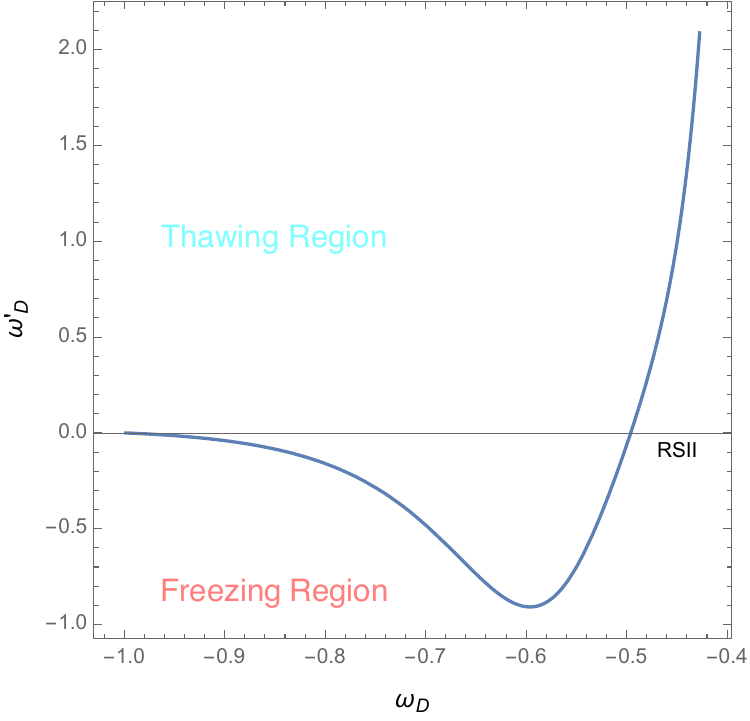}
	\caption{The $\omega^{\prime}_D-\omega_D$ diagram for RSII brane, $\Delta = 0.667 $.}
	\label{fig:frz_rs2}
\end{figure}
\begin{figure}[H]
	\centering
	\includegraphics[width = 0.8\linewidth]{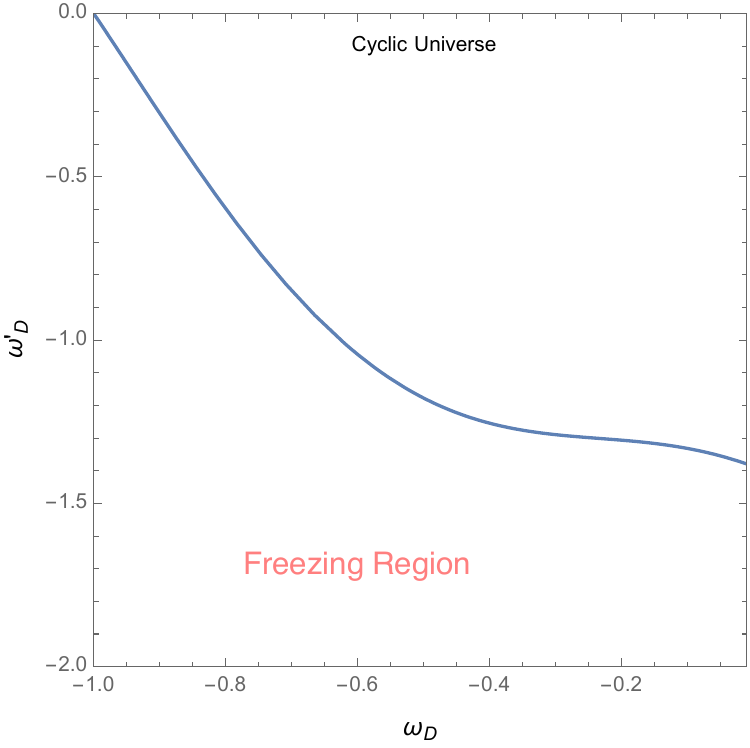}
	\caption{The $\omega^{\prime}_D-\omega_D$ diagram for the cyclic universe, $\Delta = $. $\Delta = 0.800$.}
	\label{fig:frz_cyc}
\end{figure}

Statefinder is a geometrical diagnostic tool, introduced by Sahni et. al. \cite{Sahni:2002fz}, to study the evolving nature of the dark energy. The statefinder pair, ${r, s}$, is defined as:
\begin{equation}
    r = \frac{\dddot{a}}{aH3}, \; \; \; s = \frac{r -1}{3 (q - 1/2)}.
\end{equation}
Evidently, the function $r(z)$ emerges as a natural extension of both $H(z)$ and $q(z)$. In the context of any $\Lambda$CDM model with a non-zero cosmological constant ($\Lambda$), these functions jointly assume the values of $1$ and $0$. Notably, all three models effectively converge to a $\Lambda$CDM scenario in the future, as illustrated in figures (\ref{fig:rs_dgp}) through (\ref{fig:rs_cyc}).
The DGP and RSII brane models undergo an evolutionary trajectory from a Quintessence phase ($r < 1, s > 0$) to resembling a $\Lambda$CDM model in the future. Similarly, the cyclic universe model initiates from a quintessence-like phase, eventually reducing to a $\Lambda$CDM configuration, albeit after traversing a Chaplygin gas-like phase ($r > 1, s < 0$).
At this juncture, it is pertinent to delve into the dynamical aspects of dark energy further by investigating its evolution on the $\omega^{\prime}_D - \omega_D$ plane, as detailed in \cite{caldwell2005}. The outcomes of this exploration are depicted in figures (\ref{fig:frz_dgp}) through (\ref{fig:frz_cyc}). Remarkably, both the DGP and RSII models transition from a Thawing phase ($\omega^{\prime}_D > 0, \omega_D < 1$) in the past to a Freezing phase ($\omega^{\prime}_D < 0, \omega_D < 1$). In contrast, the cyclic universe model consistently remains in the Freezing phase throughout its evolution.

\section{Classical Stability of The Model}
\label{sec:stable}
Stability of a cosmological model, at a classical level, against small perturbation can be found by calculating the adiabatic sound speed $v_s^2 = \frac{dp_D}{d\rho_D} = \frac{\dot{p}_D}{\dot{\rho}_D}$. With dark energy equation of state given by $p_D = \omega_D \rho_D$, this leads to
%---------------------------------
\begin{equation}
\label{stab}
v_{s}^{2}=\omega_{D} + \frac{\rho_{D}\dot{\omega_{D}}}{\dot{\rho_{D}}}.
\end{equation}
The models are classically stable at any stage of evolution for $v_s^2 > 0$ \cite{ghaffari2019tsallis}.

\subsection{Stability of BHDE in DGP brane-world}
\begin{figure}[H]
	\centering
	\includegraphics[width = 0.8\linewidth]{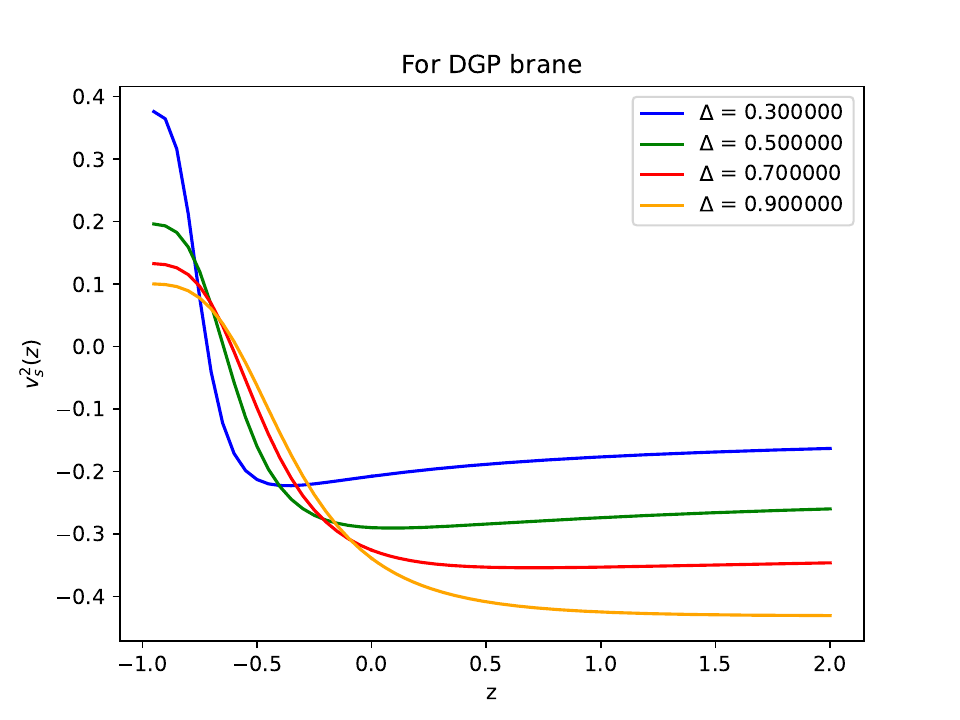}
	\caption{Evolution of adiabatic sound speed (square) in DGP brane. }
	\label{fig:vsq_dgp}
\end{figure}

Taking time derivative of eq. \eqref{eos-dgp}, 
\begin{equation}
\label{omddg}
\dot{\omega}_{D}=\frac{(\Delta -2)(\Delta- 2\epsilon(\Delta -1)\sqrt{\Omega_{rc}})}{[(\Delta -2)\Omega_{D}-2\epsilon\sqrt{\Omega_{rc}}+2]^{2}}\dot{\Omega}_{D}.
\end{equation}
Using the relation, $\frac{\rho_{D}\dot{\Omega}_{D}}{\dot{\rho_{D}}}=
\frac{\Delta\Omega_{D}}{2-\Delta}$, and \eqref{stab}
we then obtain,
\begin{equation}
\label{stadg}
v_{s}^{2}=\omega_{D} + \frac{\Delta (\Delta- 2\epsilon(\Delta -1)\sqrt{\Omega_{rc}})}{[(\Delta -2)\Omega_{D}-2\epsilon\sqrt{\Omega_{rc}}+2]^{2}}\Omega_{D}.
\end{equation}
Since the evolution of $\Omega_D$ is already know from eq. \eqref{omprime-dgp}, we can plot $v_s^2$ as a function of redshift ($z$) to study its evolution. The plot is shown in fig. (\ref{fig:vsq_dgp}). We see that BHDE in DGP brane framework is never classically stable for any value of $\Delta$ ($v_s^2 <0$). However, for any value of $\Delta$, the models attain stability at some future epoch.

\subsection{Stability of BHDE in RS II brane-world}
\begin{figure}[H]
	\centering
	\includegraphics[width = 0.8\linewidth]{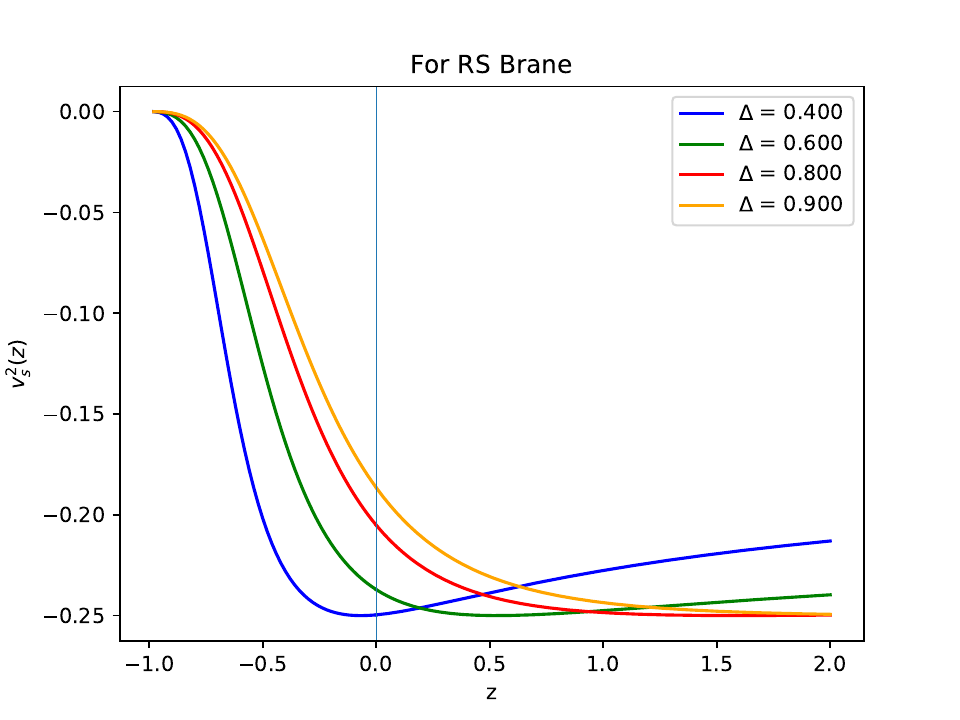}
	\caption{Evolution of adiabatic sound speed (square) in RS II brane. }
	\label{fig:vsq_rs}
\end{figure}

Taking time derivative of eq. \eqref{eosrs1} (for RS II brane, instead of subscript $D$ we are using $\Lambda$):
\begin{equation}
\label{dotomrs}
\dot{\omega}_{\Lambda}=\frac{\Delta(\Delta -2)}{2+(\Delta-2)}\dot{\Omega}_{\Lambda}.
\end{equation}
Now, using the relation, 
$\frac{\rho_{\Lambda}\dot{\Omega}_{\Lambda}}{\dot{\rho}_{\Lambda}}=
\frac{\Delta\Omega_{\Lambda}}{2-\Delta}$, and eq. \eqref{stab}
\begin{equation}
\label{stabrs}
v_{s}^{2}=\omega_{\Lambda} - \frac{\Delta^{2} \Omega_{\Lambda}}{\left(2+(\Delta-2)\Omega_{\Lambda}\right)^2}.
\end{equation}
The evolution of $v_s^2$ for RS II brane is shown in fig. (\ref{fig:vsq_rs}.

\subsection{Stability of BHDE in Cyclic Universe}
From the expression of EoS parameter in the cyclic universe (eq. \eqref{eos_cyclic}))
\begin{equation}
\label{omdcy}
\dot{\omega}_{D}= \frac{(\mathcal{A}\dot{u}+\mathcal{B}\dot{\Omega}_{D})}{\mathcal{C}},
\end{equation}
where, $\mathcal{A}=2(2-\Delta)\lbrace 2-\Omega_{D}(1+u) \rbrace-(2-\Delta)^{2}\lbrace 2-\Omega_{D}(1+u) \rbrace ^{2} - 2u(1+u)(2-\Delta)\Omega_{D}$,\;
$\mathcal{B}= -2u(1+u)^{2}(2-\Delta)$
and, \;
$\mathcal{C}=\left(2(1+u)+(\Delta-2)\lbrace 2-\Omega_{D}(1+u) \rbrace\right)^{2}$.

\begin{figure}[H]
	\centering
	\includegraphics[width = 0.8\linewidth]{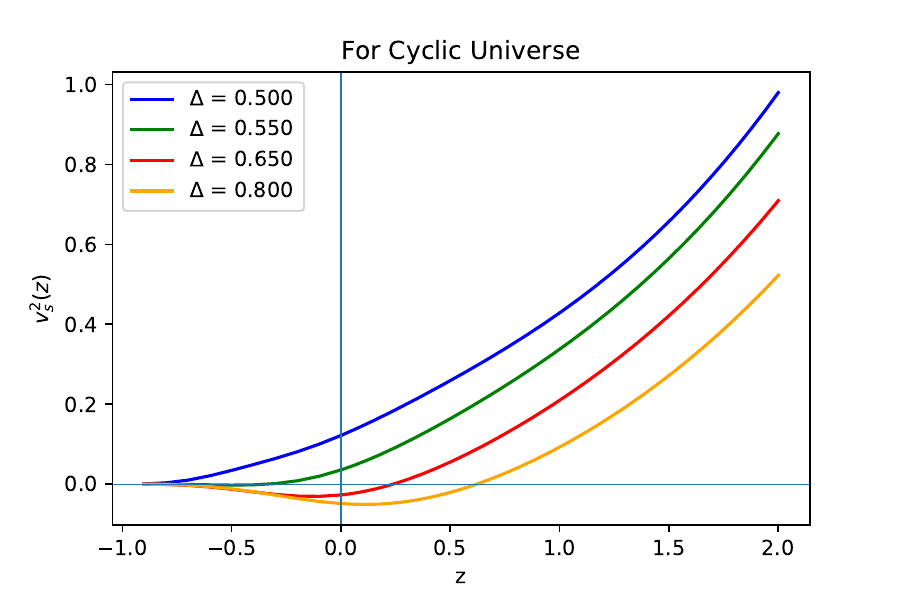}
	\caption{Evolution of adiabatic sound speed (square) in cyclic universe. }
	\label{fig:vsq_cycl}
\end{figure}

Then using the usual relation $\frac{\rho_{D}\dot{\Omega}_{D}}{\dot{\rho_{D}}}=
\frac{\Delta\Omega_{D}}{2-\Delta}$,eq. \eqref{eos_cyclic}, and eq. \eqref{stab} we can show,
$$\frac{\rho_{D}\dot{\Omega}_{D}}{\dot{\rho}_{D}}=
\frac{\Delta\Omega_{D}}{2-\Delta},~~~~ \frac{\rho_{D}\dot{u}}{\dot{\rho}_{D}}= -\frac{\omega_{D}u}{1+\omega_{D}}$$
Using the above relations we can write down $v_{s}^{2}(\Omega_{D}, u)$ explicitly. The evolution of the $v_s^2$ is depicted in fig. (\ref{fig:vsq_cycl}).
\section{Exploring the model with OHD}
\label{sec:comp}

\begin{figure}[H]
	\centering
	\includegraphics[width = 0.8\linewidth]{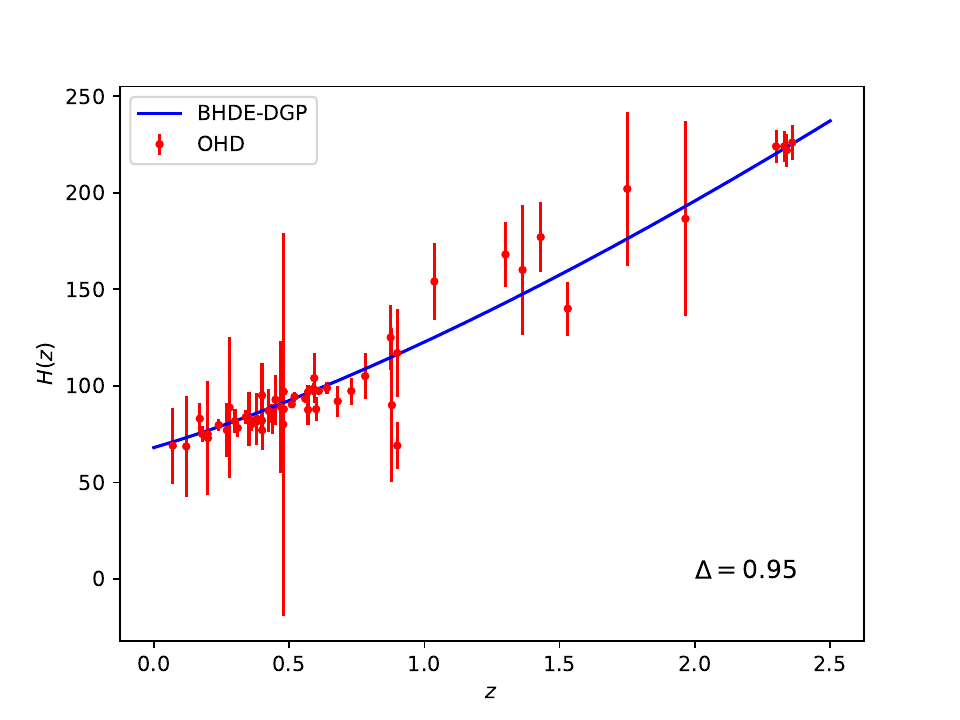}
	\caption{BHDE in DGP-Brane: Fit with OHD}
	\label{fig:bhde_dgp}
\end{figure}

\begin{figure}[H]
	\centering
	\includegraphics[width = 0.8\linewidth]{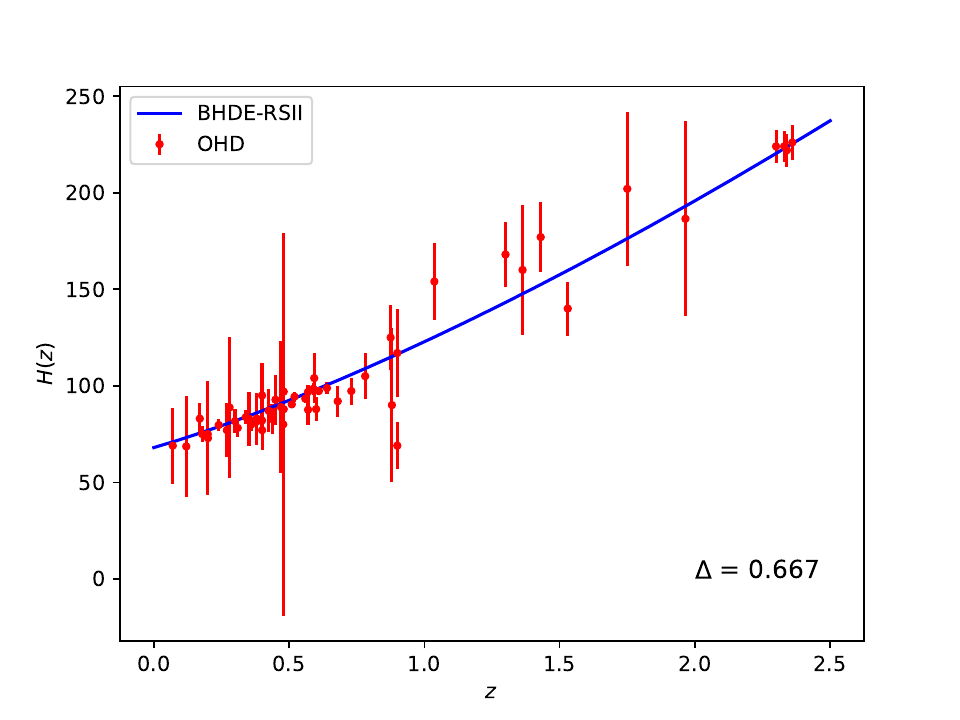}
	\caption{BHDE in RSII-Brane: Fit with OHD}
	\label{fig:bhde_rs2}
\end{figure}

\begin{figure}[H]
	\centering
	\includegraphics[width = 0.8\linewidth]{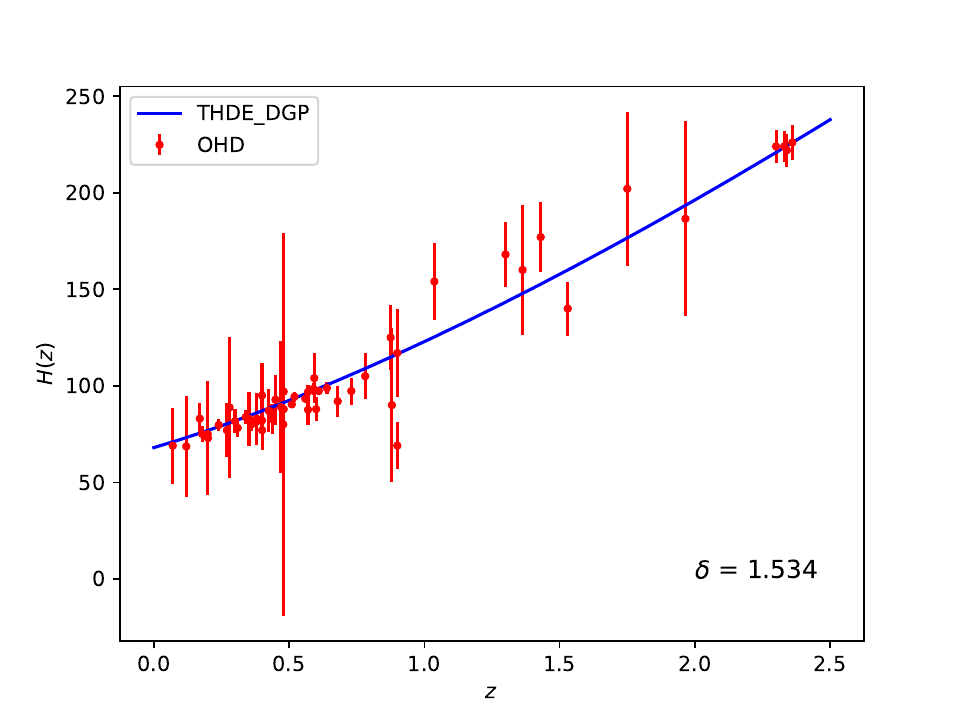}
	\caption{THDE in DGP-Brane: Fit with OHD}
	\label{fig:thde_dgp}
\end{figure}

\begin{figure}[H]
	\centering
	\includegraphics[width = 0.8\linewidth]{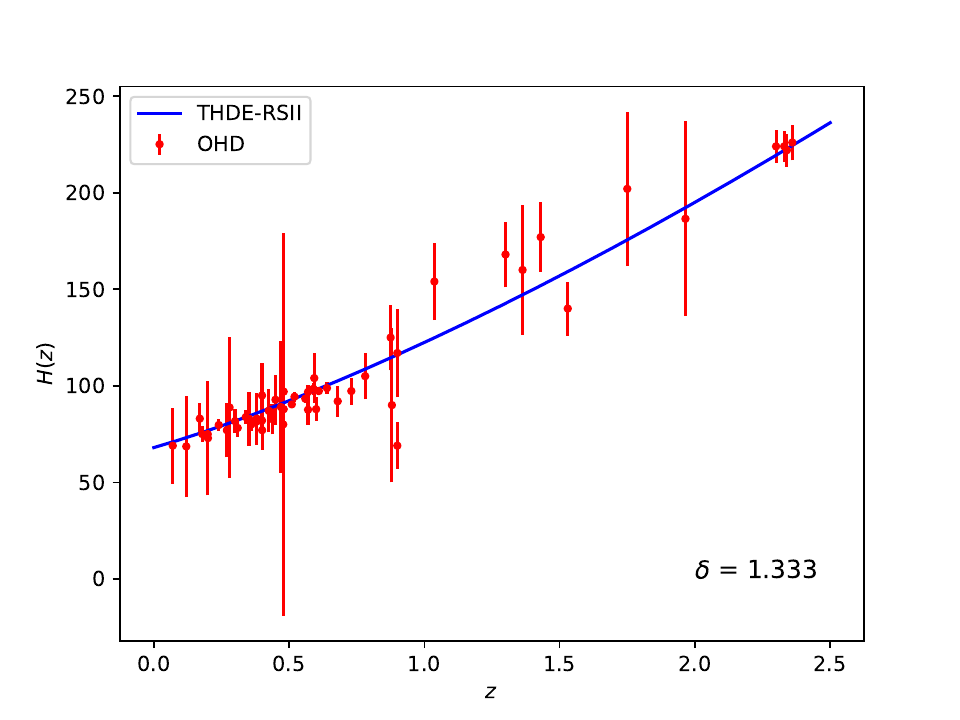}
	\caption{THDE in RSII-Brane: Fit with OHD}
	\label{fig:thde_rs2}
\end{figure}

We now compare the THDE implementation in braneworld \cite{ghaffari2019tsallis} with the present BHDE implementation using OHD. There are 57 data points altogether. $31$ of these data points are obtained from the differential ages of the galaxies and the rest from other methods (for details, see \cite{Sharov:2018yvz}). The results are summerized in table (\ref{tab:tab1}). DGP and RSII frameworks give reasonable fits for both THDE and BHDE. For THDE the best fit values of the Tsallis parameter is $\delta  = 1.54$ with the goodness of fit measure of $R^2 = 0.934$. In \cite{ghaffari2019tsallis}, different cosmological parameters are studied for different values of $\delta$, and in some cases the authors have studied the evolution with $\delta = 1.5 - \delta = 1.55$. It is difficult to comment on the other plots where $\delta > 2 $ have been used. For RSII brane though, we found the best fit value to be $\delta = 1.33$ which does not agree with the values used in \cite{ghaffari2019tsallis}.  For BHDE best fit values of the Barrow exponent (for DGP $\Delta = 0.95 $ and for RSII $\Delta = 0.667$) falls within the range suggested from theoretical consideration. Note that, OHD favours a high departure from normal HDE through the Barrow exponent value. The fitted plots are given in figs (\ref{fig:bhde_dgp}), (\ref{fig:bhde_rs2}), (\ref{fig:thde_dgp}) and (\ref{fig:thde_rs2}). However, for the cyclic universe, we could not get a good enough fit with THDE or BHDE. 

\begin{table}
\centering
\begin{tabular}{|c|c|c|}
\hline
      & BHDE ($\Delta$)                    & THDE ($\delta$)      \\        
      \hline \\
      
DGP   & $0.950 \pm 0.034$       & $1.534 \pm 0.244$ \\
\hline \\
RS II & $0.667 \pm 0.162$       & $1.333 \pm 0.180$ \\
\hline
\end{tabular}
\caption{Fitted values of BHDE and THDE parameters}
\label{tab:tab1}
\end{table}

\begin{table}
\centering
\begin{tabular}{|c|c|c|c|c|}
\hline
     Parameter & DGP & RS-II & Cyclic                    & $\Lambda$CDM      \\        
      \hline \\
      
$q$   & -0.184902       & -0.034061 & -0.627504 & -0.535\\
\hline \\
$\omega_{D}$ & -0.866138   & -0.83006 & -0.847015 & $-1.018 \pm 0.031$ \\
\hline
\end{tabular}
\caption{Present values of the deceleration parameter and the DE EoS parameter at the present universe for different brane world scenarios with BHDE and their corresponding values for the $\Lambda$CDM model.}
\label{tab:tab2}
\end{table}

%%%%%%%%%%%%%%%%%%%%%%%%%%%%%%%%%%
\section{Conclusion}
\label{sec:concl}

\begin{figure}[H]
	\centering
	\includegraphics[width = 0.8\linewidth]{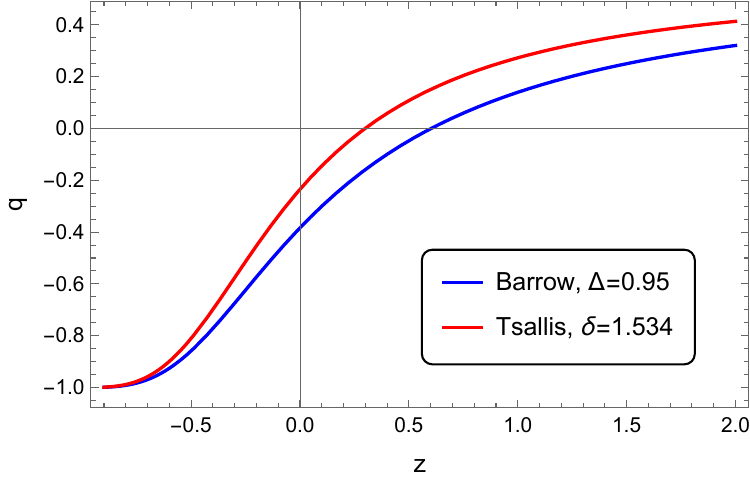}
	\caption{Evolution of deceleration parameter in DGP brane, considering for BHDE (Blue) and THDE (Red) with the best-fitted values of the model parameters obtained from OHD.}
	\label{fig:dgpcom}
\end{figure}

\begin{figure}[H]
	\centering
	\includegraphics[width = 0.8\linewidth]{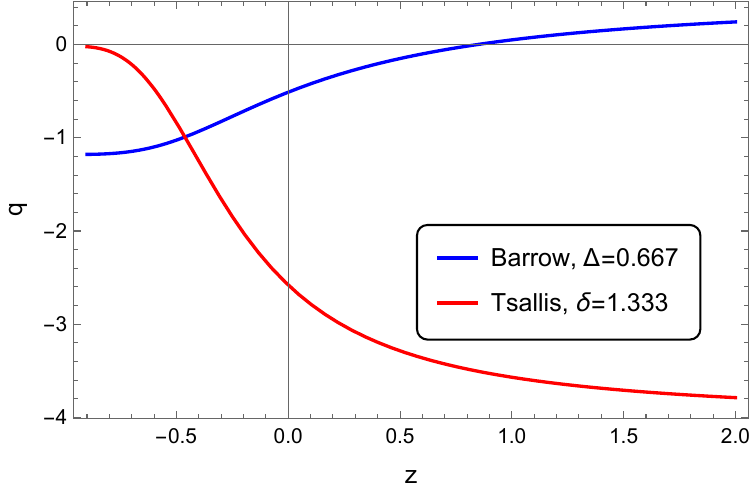}
	\caption{Evolution of deceleration parameter in Rs-II brane, considering for BHDE (Blue) and THDE (Red) with the best-fitted values of the model parameters obtained from OHD.}
	\label{fig:rs2com}
\end{figure}

\begin{figure}[H]
	\centering
	\includegraphics[width = 0.8\linewidth]{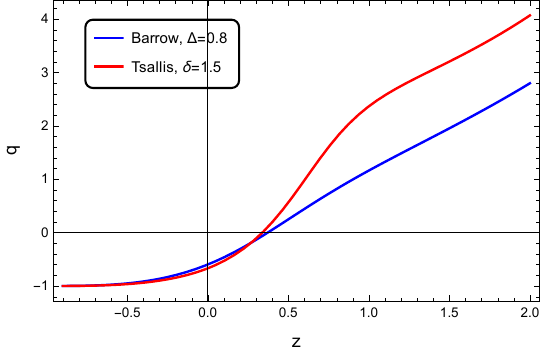}
	\caption{Evolution of deceleration parameter in Cyclic brane, considering for BHDE (Blue) and THDE (Red).}
	\label{fig:cycliccom}
\end{figure}
In this study, we have explored the Barrow Holographic Dark Energy (BHDE) description of dark energy, employing the Hubble horizon as the infrared (IR) cutoff within the context of the DGP and RS II brane world cosmologies, as well as in the evolution of a cyclic universe. Our investigation has been limited to a flat universe with no interactions between dark sectors. A comparable analysis was previously conducted with Tsallis Holographic Dark Energy (THDE) in \cite{ghaffari2019tsallis}, and intriguingly, the BHDE formulation with the Hubble horizon as an IR cutoff yields highly similar equations.
Comparison between the two models has been undertaken through the utilization of Observed Hubble Data. As demonstrated by Hsu \cite{hsu2004entropy}, the conventional Holographic Dark Energy (HDE) model fails to generate late-time acceleration when the Hubble horizon serves as the IR cutoff. To delve into this issue, we have constructed toy models of the universe, investigating the evolution of cosmological parameters and the impacts of the Barrow exponent ($\Delta$) on them.
Noteworthy observations include the convergence of all three models to the $\Lambda$CDM model in the future ($z \approx -1$) for larger $\Delta$ values. Importantly, none of the brane-world models exhibit phantom-like behavior for the considered Barrow exponent values, as depicted in figures (\ref{fig:eos_dgp}), (\ref{fig:eos_rs}), and (\ref{fig:eos_cyclic}). This behavior mirrors findings in the THDE case \cite{ghaffari2019tsallis}. The cosmological models demonstrate the capability to induce late-time acceleration, transitioning from an earlier decelerating phase, with the transition redshift ($z_t$) depending on the exponent $\Delta$. Higher $\Delta$ values push the transition redshift parameter $z_t$ further back in time.
It is notable that for smaller values of $\Delta$ (e.g., $\Delta = 0.1$), the transition occurs in the future, rendering such $\Delta$ values less realistic for cosmological modeling. This is in contrast to observations for BHDE in the General Relativity framework, where smaller $\Delta$ values are favored \cite{sardi2020}. Importantly, larger $\Delta$ values in the models developed here with BHDE indicate a greater deviation from standard HDE.
The Statefinder diagnostic (section \ref{sec:stf}) confirms the future convergence of all three models to the $\Lambda$CDM model. While RS II and DGP models transition from thawing to freezing phases, the cyclic model remains in the freezing phase throughout.
Our exploration extends to the classical stability of these models. Interestingly, the DGP brane models are inherently unstable due to a consistently negative adiabatic sound speed, as illustrated in figure (\ref{fig:vsq_dgp}). However, these models are seen to attain stability in the future. The RS II model remains unstable up to the current epoch and persists in near-future times (\ref{fig:vsq_rs}). For a cyclic universe, a critical value of $\Delta = 0.650$ is identified, below which the universe consistently maintains classical stability. Beyond this threshold, the universe undergoes a transition from stable to unstable phases at some past epoch, only to regain stability in the future (fig. \ref{fig:vsq_cycl}). However, smaller values of $\Delta$ may not be suitable for this framework to produce acceptable behavior in cosmological parameters.
It is crucial to note that our findings align with those of\cite{ghaffari2019tsallis}, revealing that neither framework is capable of producing a stable model throughout the entire evolution while maintaining acceptable behavior in cosmological parameters for any value of the Barrow exponent. The true nature of the evolution of observable cosmological parameters, even when correctly modeled, serves primarily as a foundation for constructing toy models. The construction of acceptable toy models encourages further investigations, especially in light of recent cosmological observations such as Planck 2018.
Additionally, our parameter estimation using Observed Hubble Data indicates consistency with theoretical work for BHDE. However, for THDE, the best-fit Tsallis parameter values do not align with those used in \cite{ghaffari2019tsallis} in the RS II framework. Regrettably, reasonable fits could not be obtained for the cyclic universe with either BHDE or THDE. In Table \ref{tab:tab2} we have compared the present day ($z=0$) values of the deceleration parameter ($q$) and the dark energy EoS parameter ($\omega_{D}$) for the three models along with the $\Lambda$CDM model as obtained from the PLANCK 2018 \cite{refId0}. Evolution of the deceleration parameter for THDE and BHDE are shown for all three frameworks in fig.(\ref{fig:dgpcom}), fig. (\ref{fig:rs2com}), and fig. (\ref{fig:cycliccom}). From the results, it is evident that the present day values of the cosmological parameters differ from the $\Lambda$CDM model. From the values of the DE EoS parameter it is clear that the DE remains in the quintessence regime for the present universe. These values are obtained with the best-fit values of the model parameters. Robust data analysis, capable of estimating confidence regions essential for model validation (or exclusion), thus remains an imperative avenue for future research.
Furthermore, the local nature of the classical stability analysis presented here necessitates a more rigorous study involving the examination of the growth of perturbations to conclusively comment on model stability at large scales as shown in \cite{Shtanov_2007, Cardoso2008}. This detailed analysis would also serve the dual purpose of differentiating BHDE from THDE within these frameworks. However, this aspect is beyond the scope of the present paper and will be addressed in future work.

\section*{Acknowledgement}
BCP acknowledge to DST SERB for a Project File No. CRG/2021/000183.

\bibliographystyle{iopart-num}
\bibliography{bhderef}

\end{document}